\documentstyle[aps,preprint]{revtex}
\tightenlines
\begin{document}
\draft
\title{Compressibility of Mixed-State Signals}
\author{Masato Koashi and Nobuyuki Imoto}
\address{CREST Research Team for Interacting Carrier Electronics,
School of
Advanced Sciences, \\
The Graduate University for Advanced Studies (SOKEN),
Hayama, Kanagawa, 240-0193, Japan}
\maketitle
\begin{abstract}
 We present a formula that determines 
the optimal number of qubits per message that allows  
asymptotically faithful compression of the quantum information carried by 
an ensemble of mixed states. 
The set of mixed states determines 
a decomposition of the Hilbert space into 
the redundant part and the irreducible part.
After removing the redundancy, the optimal
compression rate is shown to be given by the von Neumann entropy of 
the reduced ensemble.

\end{abstract}
 
\narrowtext

Consider a source that generates a message $i$ with 
probability $p_i$. Sequences of the messages independently
drawn from this source can be compressed into sequences of 
bits and decompressed back to the original sequences of messages.
The necessary and sufficient number of bits per message 
allowing asymptotically faithful compression and decompression 
is given by the Shannon entropy $S=-\sum_i p_i \log_2 p_i$. 
This result, called the noiseless coding 
theorem~\cite{shannon48}, is one of the core results 
of the classical information theory. 
The quantum analogue of this theorem, which will naturally form
a basis of quantum information theory, was first considered by
Schumacher \cite{schumacher95}. In this quantum data compression, 
the source emits a system in a quantum state $\rho_i$ with 
probability $p_i$, and sequences of the systems
 emitted from this source are assumed to be compressed into qubits.
It was shown~\cite{schumacher95,jozsa94,barnum96pra}
that when all $\rho_i$ are pure, the least number of qubits 
allowing asymptotically faithful recovery of the original states is 
given by the von Neumann entropy $S(\rho)=-\mbox{Tr}\rho \log_2 \rho$ 
of the density operator of the ensemble $\rho=\sum_i p_i \rho_i$.
When \{$\rho_i$\} includes mixed states, the problem is still open.
Since compression schemes applicable to the pure-state signals
can also be successfully  used for the mixed-state cases \cite{lo95}, 
the optimal compression rate $I_{\rm p}$ is bounded from above by 
the von Neumann entropy, namely, $I_{\rm p}\le S(\rho)$. 
It has also been proved 
\cite{horodecki98}
that the Levitin-Holevo function \cite{holevo73}, 
$I_{\rm LH}=S(\rho)-\sum_i p_i S(\rho_i)$, is a lower bound for 
$I_{\rm p}$, namely, $I_{\rm LH} \le I_{\rm p}$. 

The aim of this Letter is to identify the optimal compression rate
for the mixed-state  
ensemble ${\cal E}=\{p_i,\rho_i\}$.
We first introduce a function $I_{\rm R}({\cal E})$  
that is given as the von Neumann entropy of a reduced ensemble 
${\cal E}_{\rm R}=\{p_i,\sigma_i\}$. The ensemble ${\cal E}_{\rm R}$
 is derived from 
${\cal E}$ by stripping off the redundant parts.
Then we prove that $I_{\rm R}({\cal E})$ is equal to 
the optimal compression rate $I_{\rm p}$. 

The problem considered here is formulated as follows. 
Suppose that the source produces the ensemble ${\cal E}=\{p_i,\rho_i\}$,
namely, it emits a system in a quantum state $\rho_i$ with 
probability $p_i$. 
Using this source $N$ times, we obtain a
 state 
$\rho^N_\lambda\equiv \rho_{i_1}\otimes\cdots\otimes\rho_{i_N}$
acting on a Hilbert space 
${\cal H}^N\equiv {\cal H}_{1}\otimes\cdots\otimes{\cal H}_{N}$
with probability $p^N_\lambda= p_{i_1}\ldots p_{i_N}$, where $\lambda$
represents a set of indexes $\{i_1,\ldots ,i_N\}$. We assume that 
the dimension $d$ of each space ${\cal H}_{n}$ is finite.
Now ${\cal H}^N$ is given to Alice, who compresses the signal 
$\rho^N_\lambda$
into $\tilde\rho_\lambda$ acting on 
a Hilbert space ${\cal H}_{\rm C}$ with a dimension usually smaller than 
$Nd$. This process is generally written by a quantum operation 
(linear completely positive trace-preserving map) 
$\rho^N_\lambda \rightarrow 
\tilde{\rho}_\lambda=\Lambda_{\rm A} (\rho^N_\lambda)$.
The operation $\Lambda_{\rm A}$ is independent of $\lambda$
since only the systems ${\cal H}^N$ are given to Alice
and no additional information on $\lambda$ is available.
The coded signal $\tilde{\rho}_\lambda$ is passed on to 
Bob through a noiseless channel, and he decompresses the 
signal by a quantum operation  
$\tilde{\rho}_\lambda \rightarrow 
\rho^\prime_\lambda=\Lambda_{\rm B} (\tilde{\rho}_\lambda)$,
where $\rho^\prime_\lambda$ acts on ${\cal H}^N$. To measure 
the quality of the whole process 
$\rho^N_\lambda \rightarrow \rho^\prime_\lambda$, we use the 
fidelity $F$ \cite{jozsa94fid}
given by 
$F(\rho,\sigma)\equiv [\mbox{Tr}\sqrt{\rho^{1/2}\sigma\rho^{1/2}}]^2$.
The quality of a compression scheme specified by 
$(\Lambda_{\rm A}, \Lambda_{\rm B})$ for the ensemble ${\cal E}$
is given by the average fidelity
\begin{equation} 
\bar{F}\equiv\sum_\lambda p_\lambda
 F(\rho^N_\lambda, \rho^\prime_\lambda).
\end{equation}
Now, for a fixed source ${\cal E}$, consider a sequence of compression 
schemes $(\Lambda^{(N)}_{\rm A}, \Lambda^{(N)}_{\rm B})$ with increasing
$N$. When $\lim_{N\rightarrow\infty}\bar{F}=1$, the sequence gives 
asymptotically faithful compression of ${\cal E}$. Such sequences are
called {\it protocols} \cite{horodecki98}.
For a given protocol $P$, the quantity $R(P)$
characterizing the asymptotic degree of compression 
 is defined through the size of ${\cal H}_{\rm C}$ measured in 
the number of qubits, namely,
\begin{equation}
R(P)\equiv\lim_{N\rightarrow\infty}(\log_2 {\rm dim} {\cal H}_{\rm C})/N.
\label{degcomp}
\end{equation}
Then, the optimal compression rate $I_{\rm p}({\cal E})$ for the 
ensemble ${\cal E}$ is formally defined as 
\begin{equation}
I_{\rm p}({\cal E})\equiv \inf_P R(P). 
\label{passive}
\end{equation}
This means that for arbitrary 
small $\delta>0$, asymptotically faithful compression
is possible if $I_{\rm p}+\delta$ qubits per message is given, and 
it is impossible if $I_{\rm p}-\delta$ qubits per message is given.

A useful tool used for 
stripping off the redundant part in ${\cal E}$ and deriving the formula for
$I_{\rm p}({\cal E})$ below 
is the theory \cite{what} that characterizes the quantum operations
which preserves a set of states $\{\rho_i\}$ (maps $\rho_i$ to $\rho_i$)
acting on a Hilbert space ${\cal H}$.
To state the results of this theory, 
it is convenient to express quantum operations
in unitary representation, namely, by unitary operations $U$ 
acting on the combined space ${\cal H}\otimes {\cal H}_{\rm E}$, 
where ${\cal H}_{\rm E}$ represents an auxiliary system initially 
prepared in a standard pure state $\Sigma_{\rm E}$. Then, 
it was shown \cite{what} that,
given $\{\rho_i\}$,
we can find a decomposition of ${\cal H}_{\rm A}$
defined as the support of $\sum_i\rho_i$
(${\cal H}_{\rm A}$ is generally a subspace of ${\cal H}$) written as
\begin{equation}
{\cal H}_{\rm A}=\bigoplus_l
{\cal H}^{(l)}_{\rm J} \otimes {\cal H}^{(l)}_{\rm K}, 
\label{hdec}
\end{equation}
in such a way that any $U$ preserving $\{\rho_i\}$
is expressed in the following form
\begin{equation}
U(\bbox{1}_{\rm A}\otimes \Sigma_{\rm E})
=\bigoplus_l\bbox{1}^{(l)}_{\rm J}\otimes U^{(l)}_{\rm
KE}(\bbox{1}^{(l)}_{\rm K}\otimes \Sigma_{\rm E}), 
\label{main1}
\end{equation}
where $U^{(l)}_{\rm KE}$ are unitary operators acting on the combined
space ${\cal H}^{(l)}_{\rm K}\otimes {\cal H}_{\rm E}$.
Under this decomposition, $\rho_i$
is written as
\begin{equation}
\rho_i=\bigoplus_l q^{(i,l)} 
\rho^{(i,l)}_{\rm J}\otimes \rho^{(l)}_{\rm K},
\label{rhodec}
\end{equation}
where $\rho^{(i,l)}_{\rm J}$ and $\rho^{(l)}_{\rm K}$ are normalized density 
operators acting on ${\cal H}^{(l)}_{\rm J}$ and ${\cal H}^{(l)}_{\rm K}$, respectively, 
and
$q^{(i,l)}$ is the probability for the state to be  in the subspace
${\cal H}^{(l)}_{\rm J}\otimes{\cal H}^{(l)}_{\rm K}$.
 $\rho^{(l)}_{\rm K}$ is independent  of $i$, and
$\{\rho^{(1,l)}_{\rm J},\rho^{(2,l)}_{\rm J},\ldots\}$ cannot be
expressed in a simultaneously block-diagonalized form. 
An explicit procedure to obtain this particular decomposition is also 
given in \cite{what}.

The form of Eq.~(\ref{rhodec}) implies that the spaces ${\cal H}^{(l)}_{\rm K}$
are redundant in the ensemble ${\cal E}=\{p_i,\rho_i\}$. 
Consider the states $\sigma_i\equiv\bigoplus_l q^{(i,l)}
\rho^{(i,l)}_{\rm J}$ in which  the redundancy has been removed, and let
${\cal E}_{\rm R}\equiv\{p_i,\sigma_i\}$ be the corresponding ensemble. 
The von Neumann entropy of ${\cal E}_{\rm R}$ can be regarded as a function
of the ensemble ${\cal E}$, denoted as $I_{\rm R}({\cal E})$, since the 
decomposition (\ref{rhodec}) is determined by the set $\{\rho_i\}$.
What we prove below is that the optimal compression rate $I_{\rm p}({\cal E})$
is given by the function $I_{\rm R}({\cal E})$.

We begin the proof by noting that
the two ensembles ${\cal E}$ and ${\cal E}_{\rm R}$
are completely interchangeable, namely, there exist quantum operations 
$\Lambda_{\sigma\rho}$ and $\Lambda_{\rho\sigma}$ that satisfy 
$\Lambda_{\sigma\rho}(\rho^N_\lambda)=\sigma^N_\lambda$
and 
$\Lambda_{\rho\sigma}(\sigma^N_\lambda)=\rho^N_\lambda$.
If a compression scheme $(\Lambda_{\rm A}, \Lambda_{\rm B})$ for $\rho^N_\lambda$
is given, we can compose a compression scheme 
$(\Lambda_{\rm A}\Lambda_{\rho\sigma}, \Lambda_{\sigma\rho}\Lambda_{\rm B})$ 
for $\sigma^N_\lambda$. Since the fidelity does not decrease under 
a quantum operation \cite{horodecki99,barnum96},
we have the inequality
$F(\rho^N_\lambda,\Lambda_{\rm B}\Lambda_{\rm A}(\rho^N_\lambda))
\le F(\sigma^N_\lambda,
\Lambda_{\sigma\rho}\Lambda_{\rm B}
\Lambda_{\rm A}\Lambda_{\rho\sigma}(\sigma^N_\lambda))$. Hence 
the composed scheme always has a better or equal average fidelity.
This implies that if a protocol for ${\cal E}$ with an 
asymptotic degree of compression $R$
is given, we can compose a protocol for ${\cal E}_{\rm R}$ with the same 
degree $R$ \cite{horodecki98,horodecki99}. Consequently, we have
$I_{\rm p}({\cal E})\ge I_{\rm p}({\cal E}_{\rm R})$. Since a similar 
argument can be made with $\rho$ and $\sigma$ interchanged, we obtain 
the equality 
\begin{equation}
I_{\rm p}({\cal E})= I_{\rm p}({\cal E}_{\rm R}).
\label{strip}
\end{equation}

Now it is suffice to consider the cases where $\{\rho_i\}$
have no redundancy, namely, $\sigma_i=\rho_i$ and ${\cal E}_{\rm R}={\cal E}$,
and we will prove the relation $I_{\rm p}({\cal E})=S(\rho)$ in these cases. 
Since we already have the inequality $I_{\rm p}({\cal E})\le S(\rho)$,
what we need is the opposite inequality, 
$I_{\rm p}({\cal E})\ge S(\rho)$. We will give a sketch 
of the proof first. 

In a compression-decompression scheme 
$(\Lambda_{\rm A}, \Lambda_{\rm B})$, 
the state eventually evolves as $\rho^N_\lambda\rightarrow \rho^\prime_\lambda
=\Lambda(\rho^N_\lambda)$, where $\Lambda\equiv\Lambda_{\rm B}\Lambda_{\rm A}$.
In this process, the marginal state in the first system (${\cal H}_1$)
evolves from $\rho_{i_1}$ to $\mbox{Tr}_{2\ldots N}(\rho^\prime_\lambda)$.
This evolution can be regarded as a result of a quantum operation 
$\Lambda_1$, defined as 
\begin{eqnarray}
\Lambda_1 (\rho_i)&\equiv& \sum p_{i_2}\ldots p_{i_N}\mbox{Tr}_{2\ldots N}
\Lambda (\rho_i\otimes\rho_{i_2}\otimes\cdots\otimes\rho_{i_N})
\nonumber \\
&=&\mbox{Tr}_{2\ldots N}
\Lambda (\rho_i\otimes\rho\otimes\cdots\otimes\rho).
\label{deflambda1}
\end{eqnarray}
Note that $\Lambda_1$ is determined by $\Lambda$ and the {\em total} density 
operators ($\rho$) of the initial state ensembles of the other $N-1$ systems.
In a protocol,  
a scheme $(\Lambda_{\rm A}, \Lambda_{\rm B})$ for large $N$ is nearly perfect.
For this scheme, $\Lambda_1$ will almost preserve the states $\{\rho_i\}$.
The decomposition (\ref{hdec}) for $\{\rho_i\}$ satisfying $\rho_i=\sigma_i$
can be simplified as 
${\cal H}_{\rm A}=\bigoplus_l{\cal H}^{(l)}_{\rm J}$
since 
${\cal H}^{(l)}_{\rm K}$ is a one-dimensional space. Correspondingly, the 
requirement (\ref{main1}) for preserving $\{\rho_i\}$ can be written as 
\begin{equation}
U(\bbox{1}_{\rm A}\otimes \Sigma_{\rm E})=\bigoplus_l\bbox{1}^{(l)}_{\rm
J}\otimes U^{(l)}_{\rm E}\Sigma_{\rm E}, 
\label{unec}
\end{equation}
where $U^{(l)}_{\rm E}$ are unitary operators acting on 
${\cal H}_{\rm E}$.  The operation $\Lambda_1$,
which nearly preserves $\{\rho_i\}$, 
 should thus be approximately
written in the form (\ref{unec}). 
Next, take a diagonalization of the total density operator, 
$\rho=\sum_{ls} p_{l,s}|l,s\rangle\langle l,s |$, in such a way that 
for a fixed $l$, the set $\{|l,s\rangle\}$ forms a basis of  
${\cal H}^{(l)}_{\rm J}$.
Let us consider an ensemble 
${\cal E}_{\perp}\equiv\{p_{l,s}, 
\rho_{l,s}\equiv |l,s\rangle\langle l,s | \}$ composed of
orthogonal pure states. If we replace the source from ${\cal E}$
to ${\cal E}_{\perp}$ in the scheme $(\Lambda_{\rm A}, \Lambda_{\rm B})$,
the operation $\Lambda_1$ does not change because the total density operator 
 is identical for the two ensembles.
Then, the error in the transmission of 
$|l,s\rangle$ will be small since the operation of 
the form (\ref{unec}) preserves 
$\{|l,s\rangle\}$. This means that by a projection 
measurement in the basis $\{|l,s\rangle\}$, classical information 
close to $NS(\rho)$ bits can be sent through the channel 
${\cal H}_{\rm C}$. This implies 
$\log_2 {\rm dim} {\cal H}_{\rm C} \gtrsim NS(\rho)$.
Combined with the definitions (\ref{degcomp}) and (\ref{passive}),
we have  $I_{\rm p}({\cal E})\gtrsim S(\rho)$.

The strict proof is given by clarifying the meaning of 
`nearly' in the above sketch, by introducing 
several measures ($f$ and $g$ below) characterizing the 
nearness. In unitary representation,
any quantum operation for the system ${\cal H}_1$ can be 
represented by a unitary operator $U$ in $d+d^2\equiv n$ 
dimension \cite{schumacher96q}, acting on the combined 
space of ${\cal H}_1$ and an auxiliary system ${\cal H}_{\rm E}$
with dimension $d^2$.
Let us introduce two nonnegative continuous 
functions $f,g:U(n)\rightarrow R$ that measure how 
$U\in U(n)$ is close
to the form (\ref{unec}). The first one is defined as 
$f(U)\equiv 1-\sum_i p_i F(\rho_i,\Lambda_U (\rho_i))$,
where $\Lambda_U (\rho_i)\equiv {\rm Tr_E}[U(\rho_i\otimes 
\Sigma_{\rm E})U^\dagger]$.
Since $f(U)=0$ iff  $\Lambda_U (\rho_i)=\rho_i$ for all $i$,
$f^{-1}(0)$ is equal to the set of $U$ that can be expressed in 
the form (\ref{unec}). The other measure is related 
to the average error probability of the transmission 
of ${\cal E}_{\perp}$, defined as 
$p_{\rm e}\equiv 1-\sum_{l,s}p_{l,s}
\mbox{Tr}(\rho_{l,s}\Lambda_U(\rho_{l,s}))$. For later convenience,
we use the function $g(U)$ defined through 
$p_{\rm e}$, namely,
$g(U)\equiv H(p_{\rm e})+p_{\rm e}\log_2(d-1)$ with 
$H(p)\equiv -p\log_2 p-(1-p)\log_2(1-p)$.
Since the form (\ref{unec}) preserves $\{\rho_{l,s}\}$, 
$g(U)$ is zero for any $U\in f^{-1}(0)$.
An important relation between the two measures is that 
if $g$ is away from zero, $f$ must also be away from zero.
This is proved as follows.
Let us define the set 
$\bar{X}_\delta\equiv\{U|g(U)\ge \delta\}$ for arbitrary $\delta>0$.
Since $g$ is continuous, $\bar{X}_\delta$ is a closed subset of $U(n)$.
Since $U(n)$ is compact and $f$ is continuous, the image $f(\bar{X}_\delta)$
is closed in $R$. 
$\bar{X}_\delta \cap f^{-1}(0)=\emptyset$ 
implies that $0\notin f(\bar{X}_\delta)$.
Therefore, $f(\bar{X}_\delta)$ has its minimum $\eta(\delta)>0$.
This result will be used to derive the inequality (\ref{fidbound}) below.
Note that the functional dependence of $\eta$ on $\delta$ is 
determined by ${\cal E}$, and is independent of $N$.

Next, we consider the transmission of classical variable 
$\{(l,s)\}$ through the source ${\cal E}_{\perp}$ and 
the scheme $(\Lambda_{\rm A}, \Lambda_{\rm B})$.
Let $X_k(k=1,\ldots,N)$ be independent random (vector) variables 
with $\mbox{Pr}\{X_k=(l,s)\}=p_{l,s}$, and 
$X\equiv\{X_1,\ldots,X_N\}$. 
Suppose that the value of $X_k$ is encoded to the state $|l,s\rangle$
in the system 
${\cal H}_k$, the compression-decompression scheme 
is applied to combined system ${\cal H}^N$,
 and finally the state in each ${\cal H}_k$ is measured
by the projection to the basis $|l,s\rangle$, producing 
a result $Y_k$. The transmitted data is represented by 
$Y\equiv\{Y_1,\ldots,Y_N\}$. 
The quantum operation $\Lambda_k$ on each system 
${\cal H}_k$ can be written in a similar form as 
(\ref{deflambda1}). Let us take a unitary representation 
$U_k\in U(n)$ for $\Lambda_k$.
A lower bound for the mutual 
information $I(X;Y)\equiv H(X)-H(X|Y)$ 
in this example is obtained as follows.
Since $X_k$ are independent,
we have 
$H(X)=\sum_k H(X_k) =NS(\rho)$.
From the general properties of entropy,
we obtain the following inequalities \cite{cover}:
$g(U_k)\ge H(X_k|Y_k)$ (Fano's inequality),
$H(X_k|Y_k)\ge H(X_k|Y)$ (conditioning reduces entropy),
and $\sum_kH(X_k|Y)\ge H(X|Y)$ (independence bound on entropy).
Combining these, we have 
$I(X;Y)\ge NS(\rho)-\sum_k g(U_k)$.
On the other hand, $I(X;Y)$ cannot exceed the capacity 
of the channel ${\cal H}_{\rm C}$, namely, 
$\log_2 {\rm dim}{\cal H}_{\rm C}\ge I(X;Y)$.
We thus arrive at the relation 
\begin{equation}
\sum_k g(U_k)/N\ge S(\rho)-(\log_2 {\rm dim}{\cal H}_{\rm C})/N.
\label{gtohc}
\end{equation}

Now let us suppose that the number of available
qubits per message is smaller than $S(\rho)$,
namely, $(\log_2 {\rm dim}{\cal H}_{\rm C})/N=S(\rho)-\delta$
with $\delta>0$. Since the numbering of the systems 
${\cal H}_k$ is arbitrary, we can generally assume that 
$g(U_1)$ is not smaller than any other $g(U_k)$.
Then, from the relation (\ref{gtohc}) we have $g(U_1)\ge\delta$,
or equivalently, $U_1\in \bar{X}_\delta$. As shown above, 
this implies $f(U_1)=1-\sum_i p_i F(\rho_i,\Lambda_1 (\rho_i))\ge \eta(\delta)>0$.
From the properties of the fidelity function $F$, we obtain 
\begin{eqnarray}
\bar{F}&=&\sum_\lambda p^N_\lambda F(\rho^N_\lambda,
\Lambda(\rho^N_\lambda))
\nonumber \\
&\le&
\sum_\lambda p^N_\lambda F(\rho_{i_1},\mbox{Tr}_{2\ldots N}
\Lambda (\rho_{i_1}\otimes\rho_{i_2}\otimes\cdots\otimes\rho_{i_N}))
\nonumber \\
&\le&\sum_i p_i F(\rho_i,\Lambda_1 (\rho_i))\le 1-\eta(\delta)
\label{fidbound}
\end{eqnarray}
since the fidelity does not decrease under partial trace
(the first inequality) and $F(\sigma,\rho)$ is convex 
as a function of $\rho$ (the second). 
The average fidelity 
of the compression-decompression scheme never exceeds 
$1-\eta(\delta)<1$ for any $N$, where $\eta(\delta)$ is 
independent of $N$. 
This means that no protocols exist that 
satisfy $R(P)=S(\rho)-\delta$. Hence 
$I_{\rm p}({\cal E})\ge S(\rho)$. Combined with the opposite
inequality $I_{\rm p}({\cal E})\le S(\rho)$, 
we obtain $I_{\rm p}({\cal E})= S(\rho)$ for the ensemble
${\cal E}$ satisfying ${\cal E}={\cal E}_{\rm R}$.
Together with Eq.~(\ref{strip}), we obtain the formula for 
general ${\cal E}$, 
\begin{equation} 
I_{\rm p}({\cal E})=I_{\rm R}({\cal E}),
\label{main}
\end{equation}
which is the main result of this Letter. For convenience,
we repeat the definition of the function $I_{\rm R}({\cal E})$:
From ${\cal E}=\{p_i,\rho_i\}$, determine 
${\cal E}_{\rm R}=\{p_i,\sigma_i\}$ through the decomposition 
(\ref{rhodec}). Then, $I_{\rm R}=S(\sigma)$ with 
$\sigma=\sum_i p_i\sigma_i$.

The protocols we considered above is asymptotically reversible, 
namely, Bob is required to 
asymptotically reproduce everything that was given to Alice.
Bob can thus compress the reproduced signals again with the
same degree of compression. This class of protocols is called 
{\em blind} protocols, and the obtained bound $I_{\rm p}$ is called
{\em passive information} \cite{horodecki98,horodecki99}. 
In another scenario, not only 
the system ${\cal H}^N$ but also the identity of the state $\rho^N_\lambda$,
namely, the index $\lambda$ is disclosed to Alice.
Bob still has to decompress the signal without additional knowledge of $\lambda$.
This class of protocols is called 
{\em visible} protocols, and the corresponding optimal compression 
rate is called
{\em effective information} $I_{\rm eff}$ \cite{horodecki98,horodecki99}.
This scheme is irreversible and cannot be repeated, but the 
compression rate $I_{\rm eff}$ may be better than $I_{\rm p}$. 
The difference $I_{\rm d}\equiv I_{\rm p}-I_{\rm eff}$
is called {\em information defect}. 
For an ensemble of pure states, it was shown that the information 
defect is zero \cite{barnum96pra}. While the identity of $I_{\rm eff}$
is still an open question, the derived form of $I_{\rm p}$ assures 
the presence of nonzero information defect for an ensemble of mixed states,
which can be shown as follows. 
In the second scenario, Alice can compress the classical 
value $\lambda$ into the length of Shannon entropy,
and send it directly to Bob. This indicates 
$I_{\rm eff}\le -\sum_i p_i\log_2 p_i$. For example, if $p_1=p_2=1/2$,
$I_{\rm eff}\le 1$. On the other hand, by allowing the dimension $d$ large,
we can find examples of $\rho_1$ and $\rho_2$ with arbitrarily large $I_{\rm p}$,
according to the result (\ref{main}).

Finally, we would like to raise several problems which is worthy
of future investigation.
What we have proved in this Letter corresponds to the so-called 
the weak converse of Shannon's noiseless coding theorem, namely, 
if $I_{\rm p}-\delta$ qubits are available per system, 
the fidelity cannot reach unity in $N\rightarrow \infty$.
For classical or pure-state ensembles, the strong converse holds,
namely, the fidelity goes to zero when $N\rightarrow \infty$. 
Whether this statement holds for mixed-state cases or not is an important
open question. In the proof of the main result, we utilized the 
observation that any protocols for an mixed-state ensemble ${\cal E}$
with no redundancy (${\cal E}={\cal E}_{\rm R}$) can be used 
to transmit the `purified' ensemble ${\cal E}_\perp$ with errors asymptotically
negligible {\em per message}. 
The requirement ($\bar{F}\rightarrow 1$) for the compression protocols 
for ${\cal E}_\perp$ is more stringent, namely, the total errors for 
the whole $N$ messages must be negligible. 
Whether the protocols for ${\cal E}$
always works as compression protocols for ${\cal E}_\perp$ or not is another 
open question.

In summary, we derived the formula for the optimal compression rate
(passive information) 
for a general mixed-state ensemble $\{p_i, \rho_i\}$. This will give 
a measure of how much information is stored in the ensemble of 
quantum states in terms of qubits. We have also shown the
presence of nonzero information defect, namely, there are cases where 
knowing the identity of states gives more efficient compression.

This work was supported by a Grant-in-Aid for Encouragement of Young
Scientists (Grant No.~12740243) and a Grant-in-Aid for Scientific 
Research (B) (Grant No.~12440111)
by Japan Society of the Promotion of
Science.

\end{document}